\documentclass[twocolumn,showpacs,aps,pra]{revtex4}
\usepackage{epsfig}

\begin{document}

\title{Measurement of the Casimir-Polder force through center-of-mass oscillations of a Bose-Einstein condensate}
\author{D.~M. Harber, J.~M. Obrecht, J.~M. McGuirk,$^*$ and E.~A. Cornell$^\dag$}
\affiliation{JILA, National Institute of Standards and Technology and University of Colorado Department of Physics, \\
University of Colorado, Boulder, Colorado 80309-0440}

\date{\today}

\begin{abstract}
We have performed a measurement of the Casimir-Polder force using
a magnetically trapped $^{87}$Rb Bose-Einstein condensate.  By
detecting perturbations of the frequency of center-of-mass
oscillations of the condensate perpendicular to the surface, we
are able to detect this force at a distance $\sim$5 $\mu\mbox{m}$,
significantly farther than has been previously achieved, and at a
precision approaching that needed to detect the modification due
to thermal radiation. Additionally, this technique provides a
limit for the presence of non-Newtonian gravity forces in the
$\sim$1 $\mu\mbox{m}$ range.
\end{abstract}

 \maketitle

\section{Introduction}

Interest in the Casimir-Polder~\cite{casimir1948} force, the
attractive QED force between an atom and a surface, and the
closely related Casimir force, the attractive QED force between
two surfaces, has blossomed in recent years following the
breakthrough experiments of Sukenik \textit{et
al.}~\cite{hinds1993} and Lamoreaux~\cite{lamoreaux1997}.
Additionally, the tremendous experimental progress in both
ultracold atomic systems and microelectromechanical systems
(MEMS's), has pushed both fields towards precise work very close
to surfaces---regimes where Casimir-type effects become important.

To the present, experiments have identified the crossover in
behavior between the van der Waals-London and Casimir-Polder
regimes, which occurs at an atom-surface separation of $\sim$0.1
$\mu\mbox{m}$ for $^{87}$Rb. Inside of this crossover, the van der
Waals-London regime, the potential scales as $1/d^{3}$, where $d$
is the distance between the atom and surface.  Outside of this
crossover, the Casimir-Polder regime, the potential scales as
$1/d^{4}$.  It has been predicted that the presence of thermal
radiation from the surface and surroundings will modify the
behavior of this force.  This modification is predicted to occur
at even greater atom-surface separations, i.e. $\sim$7
$\mu\mbox{m}$ at 300 K. In this large-separation regime, hereafter
referred to as the thermal regime, the potential scales as
$T/d^{3}$, where $T$ is the temperature of the thermal blackbody.

The experiment of Sukenik \textit{et al.}~\cite{hinds1993} was the
first to clearly measure the crossover from the van der
Waals-London to the Casimir-Polder regime.  A number of
experiments followed that used ultracold atoms to detect the
presence of the Casimir-Polder
force~\cite{aspect1996,shimizu2001,dekieviet2003,vuletic2004,ketterle2004,shimizu2005};
however, none have approached the sensitivity at large distances
required to detect a crossover to the thermal regime.

Perhaps the most obvious technique for measurements of surface
forces using ultracold atoms is interferometry in which atoms take
separate spatial paths.  This sort of atom interferometry has
proved difficult, but some groups have now started to make headway
in this direction~\cite{inguscio2004,cornell2005,ketterle2005}.

In this experiment, the effects of the surface potential on the
\textit{mechanical} motions of a Bose-Einstein condensate are
studied. The attractive surface force distorts the trapping
potential and thus manifests itself in a number of ways. First,
the center-of-mass position of the atoms changes, but only by
$\sim$10 nm for our 228 Hz trapping potential when the condensate
is several microns from the surface.  Position deviations on this
order are well below our experimental sensitivity; thus, detection
of the Casimir-Polder force in this manner is unfeasible.  Second,
the collective oscillation frequencies of the condensate change.
Of these, the center-of-mass oscillation, or dipole oscillation,
is perhaps the most robust because it is very long lived and its
frequency is independent of intracondensate interactions.

\begin{figure}
\leavevmode
\epsfxsize=3.375in 
\epsffile{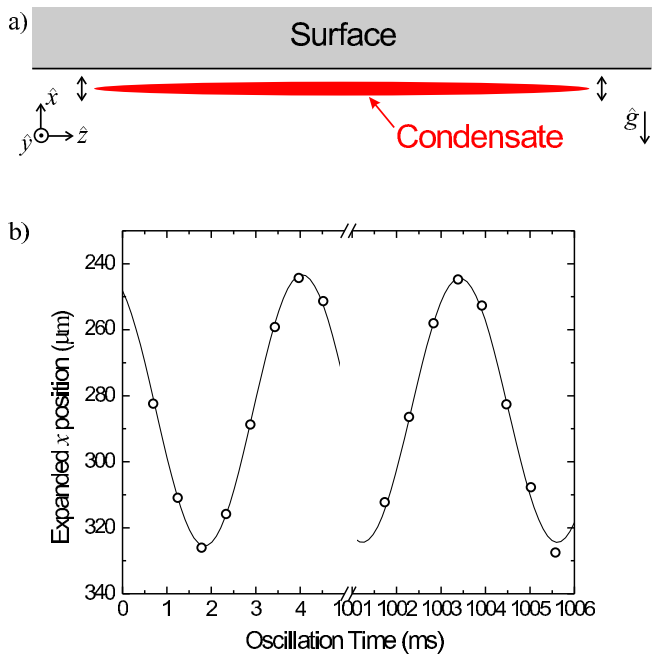} \caption{\label{fig:diagram} (Color
online) (a) Diagram, to scale, illustrating the aspect ratio of
the condensate and typical oscillation position relative to the
surface.  The coordinate axis orientation and the direction of
gravity are also indicated.  (b) Typical data showing the radial
dipole oscillation after expansion away from the surface.}
\end{figure}

In the most simple approximation, the normalized dipole
oscillation frequency shift in the $\hat{x}$ direction [see
Fig.~\ref{fig:diagram}(a)], hereafter referred to as $\gamma_x$,
can be written as

\begin{equation}
\gamma_x \equiv \frac{\omega_x-\omega'_x}{\omega_x} \simeq - \frac{1}{2 \omega_x^2 m}\partial^2_x U^*,
\end{equation}

\noindent where $\omega_x$ is the unperturbed trap frequency in
the $\hat{x}$ direction, $\omega'_x$ is the perturbed trap
frequency, $m$ is the mass of $^{87}$Rb, $\partial^2_x$ is the
second partial derivative with respect to $x$, and $U^*$ is the
potential experienced by the atoms due to the surface.  Thus the
dipole oscillation frequency is sensitive to the second derivative
of the potential, or to force {\it gradients}.

A detailed theoretical analysis of this system was performed by
Antezza \textit{et al.}~\cite{stringari2004}.  This analysis
includes a careful calculation of the Casimir-Polder force from a
dielectric surface, the modification to the Casimir-Polder force
in the thermal regime, and the expected normalized dipole
frequency shift $\gamma_x$, taking into account the finite width
of the condensate and the finite oscillation amplitude.  The
result of this analysis is that for experimentally plausible
conditions, the expected values of $\gamma_x$ are on the order of
$10^{-4}$, well within experimental precision.

\section{Experiment}

We briefly review the apparatus for generating condensates and
measuring surface forces, as a more detailed description of the
apparatus used to produce the condensate can be found
in~\cite{lewandowski2003} and the technology and techniques for
atom-surface measurements are described in detail
in~\cite{harber2003,mcguirk2004}.  At the end of evaporation,
nearly pure condensates (the fraction of atoms in the condensate
$\geq0.8$) of $1.4\times 10^5$ magnetically trapped $^{87}$Rb
atoms are created in the $|F=1,m_{F}=-1\rangle$ ground state.  In
our Ioffe-Pritchard-type magnetic trap, with trapping frequencies
of 6.4 Hz in the axial direction ($\hat{z}$) and 228 Hz in the
radial directions ($\hat{x}$ and $\hat{y}$), this corresponds to
condensate Thomas-Fermi radii of 85.9 $\mu\mbox{m}$ and 2.40
$\mu\mbox{m}$ in the axial and radial directions, respectively.
See Fig.~\ref{fig:diagram}(a) for the coordinate definitions and
orientations of the surface and condensate in the experiment.

The surfaces for study are located $\sim$1 mm above ($+\hat{x}$
direction) where evaporation occurs.  To position the condensate
near the surface, a vertical ($\hat{x}$ direction) magnetic field
is applied.  This uniform magnetic field acts to displace the
magnetic minimum of the trapping field.  By applying a carefully
controlled field ramp, we are able to move the atoms arbitrarily
close to the surface without exciting mechanical oscillations of
the condensate, and the condensate can be held there for many
seconds.

To measure the distance between the condensate and the surface, we
use an absorption imaging technique described
in~\cite{schmied2003,harber2003} where we illuminate the atoms
with a beam perpendicular to the long axis of the condensate. This
beam impinges on the surface with a slight grazing incidence angle
of $\sim$$2.4^{\circ}$ such that when the condensate is within
$\sim$100 $\mu\mbox{m}$ of the surface, both a direct absorption
image and a reflected absorption image of the condensate appear.
Measuring the distance between these images allows us to determine
the distance between the condensate and surface.  To calibrate the
magnetic field necessary to position the condensate a given
distance from the surface, a series of images are taken where we
push the atoms to a range of distances $\sim$20--60 $\mu\mbox{m}$
from the surface. The condensate-surface separations in these
images are measured and then used for calibration of the magnetic
field used to push the atoms.

To allow measurement of surface forces at different surface
locations, the magnetic trap can be moved independently of the
surface in the $\hat{y}$ and $\hat{z}$ directions.  Since the
condensate only interacts with a $\sim$$200 \times 10$
$\mu\mbox{m}$ region of the surface, we can translate the trap to
measure surface forces at many different locations on our
$5\times8$ mm surfaces.  Finally, we can adjust the angle of the
$\hat{z}$ trap axis to be parallel with respect to the surfaces.
Using the surface reflection images, we have verified that the
deviation from parallel is $\leq0.25^{\circ}$.

To excite a condensate dipole oscillation in the $\hat{x}$
direction, we apply an oscillating magnetic field of the form

\begin{equation}
B_x(t) \propto e^{-(t-t_0)^2/\tau^2} \mbox{cos}(\omega_x t),
\end{equation}

\noindent where $\tau$ is the time width of the pulse (10 ms in
this experiment) and $t_0$ is the time of the peak of the pulse.
In frequency space, this excitation is centered on the radial trap
frequency $\omega_x$ and contains no dc or high-frequency
components; this prevents excitation of unwanted internal
condensate modes.  Similarly, dipole oscillations can be excited
in the $\hat{y}$ and $\hat{z}$ directions.

Expansion of the oscillating condensate is accomplished by a
microwave adiabatic rapid passage to the $|F=2,m_{F}=-2\rangle$
state, which is antitrapped, followed by $\sim$5 ms of rapid
antitrapped expansion~\cite{lewandowski2003}.  The antitrapped
expansion acts to push atoms away from the magnetic minimum, and
because of gravitational sag, the condensate begins the expansion
below the magnetic minimum, so the condensate is pushed away from
the surface during expansion.  Additionally, the antitrapped
expansion acts to amplify the radial dipole oscillation amplitude
by approximately 20-fold, permitting straightforward measurement
of the oscillation in expansion.  For example, see
Fig.~\ref{fig:diagram}(b).  Finally, the condensate is
simultaneously imaged through absorption along both the $\hat{y}$
and $\hat{z}$ directions, allowing us to monitor the position of
the condensate in all three directions.

The typical experiment is performed as follows.  First, a surface
calibration set is taken to determine the magnetic field necessary
to position the condensate the desired distance from the surface.
Second, a vertical oscillation data set is taken at the desired
trap-center to surface distance $d$, typically 6--12
$\mu\mbox{m}$.  Interspersed with these data are vertical
oscillation data taken at $d_0$, the distance we use to obtain the
normalization frequency $\omega_x$.  Data points and normalization
points were randomly alternated during the course of the data set
in order to prevent trap frequency drift from affecting our
measurement.  For this experiment $d_0 = 15$ $\mu\mbox{m}$.  A
distance of 15 $\mu\mbox{m}$ is far away enough such that surface
forces will not affect the frequency; the normalized dipole
frequency shift from the Casimir-Polder force is less than
$10^{-6}$ at this distance.  By comparing the frequency measured
at $d$ to that measured at $d_0$, we obtain $\gamma_x$ at the
particular condensate-surface separation $d$.  Last, a second
surface calibration set is taken to determine the
condensate-surface distance drift over the course of the data set,
typically $\ll 1$ $\mu\mbox{m}$.

\section{Systematics}

Rejection of the presence of spurious forces on the condensate
caused by surface-based electric and magnetic fields is critical
to the interpretation of our results. Our most powerful test for
spurious forces is provided by the elongated geometry of our
condensate. The mean condensate-surface separation for our closest
measurements, $\sim$6 $\mu\mbox{m}$, is significantly smaller than
the axial extent of the condensate, $\sim$170 $\mu\mbox{m}$ [see
Fig.~\ref{fig:diagram}(a)]. Thus a spatially inhomogeneous force,
due to localized electric or magnetic surface contamination, would
likely affect only part of the condensate.  When oscillating in
this inhomogeneous potential, the condensate behaves more like a
string than a stiff bar and thus will oscillate at different
radial frequencies along its axial extent. Using a technique fully
described in~\cite{mcguirk2004}, we analyze images of the
oscillating condensate and obtain $\gamma_x(z)$ axially along the
center $\sim$120 $\mu\mbox{m}$ of the condensate, in addition to
$\gamma_x^{cm}$ for the center of mass of the condensate.

To investigate the presence of a spatially inhomogeneous force, we
examine $\gamma_x(z)$ across the condensate.  We first define the
standard deviation of $\gamma_x(z)$ along the axial extent of the
condensate to be $\delta_\gamma$.  If $\delta_\gamma$ is greater
than a predefined threshold value \cite{threshold}, then we
surmise that there is a statistically significant spatially
inhomogeneous force acting on the condensate and thus move to a
new surface location. On the other hand, if $\delta_\gamma$ is
less than the threshold value, then we take $\delta_\gamma$ to be
our systematic limit on spatially inhomogeneous forces experienced
by the condensate.

Spatially uniform spurious shifts along the extent of the
condensate are less likely but must also be accounted for.  A
completely uniform surface charge, or magnetization, will by
symmetry not generate a force.  The remaining possible cause of
spurious uniform forces is then stripes of surface contaminations
collinear with the axis of the condensate~\cite{adsorbates}.  To
test for this possibility, we perform measurements at multiple
surface locations, as well as a series of measurements to test for
the presence of magnetic and electric fields.

An atom in an electric field will experience an energy shift
according to $U_E = -(\alpha_{0}/2)E^{2}$, where $\alpha_{0}$ is
the ground-state dc polarizability and $E$ is the electric field
magnitude.  Thus, to first approximation, the normalized frequency
shift caused by an electric field can be written as

\begin{equation}
\gamma_x \propto - \partial^2_x U_E =
\frac{\alpha_0}{2}\partial^2_x\left[(E^*_x)^2 + (E^*_y)^2 +
(E^*_z)^2\right].
\end{equation}

\noindent Our goal is then to determine the $x$ dependence of the
surface electric fields $E^*_x$, $E^*_y$, and $E^*_z$; from these
we can obtain an estimate of $\gamma_x$. In our previous
work~\cite{mcguirk2004}, we applied a uniform dc external electric
field $E^{ext}_x$, and by measuring a change in $\gamma_x$ as
$E^{ext}_x$ was varied $\pm100$~V/cm, we were able to obtain
$\partial^2_xE^*_x(x)$. For the current work, we are instead using
a technique that allows us to measure $\partial_xE^*_x(x)$ as well
as $\partial_xE^*_y(x)$ and $\partial_xE^*_z(x)$.

If rather than apply a dc external electric field we apply an ac
external electric field $E^{ext}_x cos(\omega t)$, where
$E^{ext}\gg E^*$, and we invoke $\vec{\nabla} \times \vec{E}
\simeq 0$~\cite{bdot}, then the forces on the atoms are

\begin{eqnarray}
F_x(t) & \simeq & \alpha_0 E^{ext}_x cos(\omega t) \partial_xE^*_x, \\
F_y(t) & \simeq & \alpha_0 E^{ext}_x cos(\omega t) \partial_xE^*_y, \\
F_z(t) & \simeq & \alpha_0 E^{ext}_x cos(\omega t) \partial_xE^*_z.
\end{eqnarray}

If $\omega$ is set to $\omega_x$, $\omega_y$, or $\omega_z$ and
$\partial_xE^*_{x,y,z}$ is nonzero, then the oscillating electric
field will resonantly drive a dipole oscillation.  This allows the
measurement of very small electric forces.  By keeping the drive
time short, 50--100 ms in our experiment, compared to the
oscillation damping rate, typically 1--10 s, and the inverse of
the drive detuning, typically $<0.5$ Hz, then the system reduces
to that of an undamped, resonantly driven oscillator. In this
case, the oscillation amplitude linearly grows as

\begin{equation}
\dot{A} = \frac{F_0}{2 \omega m},
\end{equation}

\noindent where $F_0$ is the amplitude of the driving force and
$\dot{A}$ is the rate of growth of the amplitude. By positioning
the condensate at a given distance from the surface and driving a
large external electric field ($\sim$100~V/cm) at the $\hat{i}$
trap frequency, we are able to detect $\partial_xE_i^*$, where
$\hat{i}$ represents the $\hat{x}$, $\hat{y}$, or $\hat{z}$
direction.  This measurement is performed at multiple distances
from the surface, and $\partial_xE_i^*(x)$ is fit to a form
$\partial_xE_i^*(x) = -p_i C_i/x^{p_i+1}$.  This corresponds to an
electric field of the form $E_i^*(x) = C_i/x^{p_i}$.  When fitting
this data for $C_i$ we vary the exponent $p_i$ between 0.20,
corresponding to a very broad collinear surface charge or dipole
distribution, and 2.0, corresponding to a line of dipoles.
Finally, using this range of powers, the power-law fit leading to
the largest systematic error is used.

Although less prevalent, evidence of magnetic surface contaminants
can be seen on our insulating surface (likely remnants from
surface polishing).  Our trapping potential itself is magnetic, so
we cannot rely on techniques similar to those used for the
detection of electric fields. Instead, we carefully examine the
magnetic trapping potential itself. A spurious magnetic field
cannot exclusively modify $\omega_x$; rather, the spurious field
will manifest itself as a modification of multiple trapping
frequencies, or anomalous center-of-mass displacements.

The trap frequencies in the three directions can be expressed as

\begin{eqnarray}
\omega_x  = \omega_y & = & \sqrt{\frac{\mu_B m_F g_F }{m}} \frac{\eta}{\sqrt{B_0}},\\
\omega_z & = & \sqrt{\frac{\mu_B m_F g_F }{m}} \sqrt{\beta},
\end{eqnarray}

\noindent where $\mu_B$ is the Bohr magneton, $g_F$ is the
Land\'{e} $g$ factor, $\eta$ is the linear magnetic field gradient
in the $\hat{x}$ and $\hat{y}$ directions~\cite{xygradient}, $B_0$
is the bias field, and $\beta$ is the magnetic field curvature in
the $\hat{z}$ direction [$B_z(z) = B_0 + \beta/2~z^2$]. Adding an
additional spurious magnetic field $B^*$ and expanding to first
order in $B^*$, the normalized frequency shift $\gamma_{x,y,z}$
induced by the spurious magnetic field can be written as

\begin{eqnarray}
\gamma_x & = & \frac{B^*_z}{2 B_0} - \frac{\partial_xB^*_x}{\eta} - \frac{B_0 \partial_x^2B^*_z}{2\eta^2},\\
\gamma_y & = & \frac{B^*_z}{2 B_0} + \frac{\partial_yB^*_y}{\eta} - \frac{B_0 \partial_y^2B^*_z}{2\eta^2},\\
\gamma_z & = & - \frac{\partial_z^2B^*_z}{2\beta}.
\end{eqnarray}

\noindent Invoking $\vec{\nabla} \cdot \vec{B}^* = 0$ and
$\vec{\nabla} \times \vec{B}^* \simeq 0$~\cite{edot}, we obtain an
expression for the systematic uncertainty in the normalized
frequency shift caused by a spurious magnetic field, $\delta
\gamma_x$, as

\begin{equation}
\delta \gamma_x \simeq \sqrt{(\delta \gamma_y)^2 + \frac{ B_0^2 \beta^2 }{\eta^4} (\delta \gamma_z)^2
+ \frac{\beta^2}{\eta^2} (\delta z_{cm})^2},
\end{equation}

\noindent where we have introduced $\delta \gamma_y$ and $\delta
\gamma_z$, which are the measured systematic limits on deviations
from zero of $\gamma_y$ and $\gamma_z$ as the condensate nears the
surface, and $\delta z_{cm}$, which is the uncertainty in the
movement of the center of mass of the condensate in the $\hat{z}$
direction from its equilibrium position.

This technique can be summarized as follows.  If the only force on
the condensate is the Casimir-Polder force, then $\gamma_y$,
$\gamma_z$, and $z_{cm}$ will not change as the condensate is
moved near the surface.  Therefore, if $\gamma_y$, $\gamma_z$, and
$z_{cm}$ display no statistically significant deviation, then we
have verified that magnetic forces from the surface are not
present at measurable levels.  The uncertainty in these terms,
$\delta \gamma_y$, $\delta \gamma_z$, and $\delta z_{cm}$, then
allow us to obtain a systematic limit on the presence of magnetic
forces.

This technique, where we only consider uncertainties of fields, is
not applicable in the case of the electrical systematics because
typically a nonzero electric field is detected.  Therefore, the
perturbation of the measured electric field, rather than the
uncertainty in the presence of an electric field, dominates the
electrical systematics.

Finally, we acknowledge that in certain situations, a magnetic
field could cause a change in $\gamma_x$ and no change in
$\gamma_y$--for instance, if $B^*_z/(2 B_0) = -\partial_yB^*_y /
\eta$.  Unlikely situations such as this cannot be categorically
excluded with our current analysis technique; however, by
performing measurements at multiple surface locations and
verifying that these measurements agree, we can reduce the
possibility that this sort of unusual cancellation could disturb
our measurement.

\begin{table}
\begin{center}
\label{tbl:errors} \caption[Error Budget]{A summary of our errors
in $\gamma_x$.  The $1\sigma$ error bars in Fig.~\ref{fig:data}
represent a combination of statistical and systematic errors.  The
relative contributions of the various sources of statistical and
systematic error were evaluated separately for each point.  Values
for the errors for the worst point and for a typical point are
tabulated under "Maximum" and "Typical" below.}
    \begin{tabular}{l c c}
      \hline \hline
          Error & Maximum & Typical \\
          source & ($10^{-5}$) & ($10^{-5}$) \\ \hline
          Statistical & 8.3 & 4.0 \\
          Spatial inhomogeneity & 4.6 & 2.5 \\
          Uniform magnetic & 2.9 & 2.2 \\
          Uniform electric & 4.1 & 0.41 \\
          Normalization & 0.18 & 0.14 \\ \hline
          Total & - & 5.2 \\ \hline \hline
    \end{tabular}
\end{center}
\end{table}

\section{Results}

\begin{figure}
\leavevmode
\epsfxsize=3.05in 
\epsffile{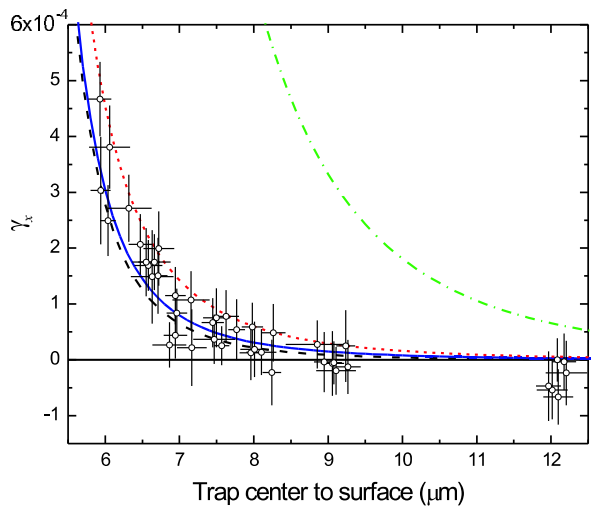} \caption{\label{fig:data} (Color
online) Normalized frequency shift data from the fused silica
surface (dc dielectric constant = 3.83~\cite{jarvis1977}).  Each
data point represents a single measurement of $\gamma_x$ (no data
averaging was performed).  These data were taken at two locations
spaced 300 $\mu\mbox{m}$ apart.  Vertical error bars include the
statistical and systematic errors detailed in Table~I.  Horizontal
error bars include statistical uncertainty, surface drift, and the
effects of the uncertainty in the image magnification.  For this
data set the mean oscillation amplitude, including the small
effects of damping, is 2.06 $\mu\mbox{m}$.  The mean Thomas-Fermi
radius in the $\hat{x}$ direction is 2.40 $\mu\mbox{m}$ for this
data. Theory lines, calculated using the theory from Antezza
\textit{et al.}~\cite{stringari2004,static-approx}, indicate $T$ =
0 K (dashed, black line), $T$ = 300 K (solid, blue line), and $T$
= 600 K (dotted, red line). Additionally we include the
extrapolation of the van der Waals-London $1/d^3$ potential to
these distances (dash-dotted, green line).}
\end{figure}

The surfaces with which these experiments were performed were
$\sim8\times5\times2$ mm$^3$ pieces of UV-grade fused silica and
sapphire polished to $\sim$0.5 nm surface roughness.  Conducting
surfaces would in some way be preferable, largely because they are
less susceptible to electric fields caused by surface charge.
Unfortunately, alkali-metal atoms, when adsorbed on a conducting
surface, generate a significant electric dipole
field~\cite{mcguirk2004}.  Preventing any atoms from striking the
surface during a measurement is unfeasible, so is seems that
dielectric surfaces, despite the possible presence of surface
charges, are preferable in this case.

The dielectric surfaces studied include (1) UV-grade fused silica
prepared by a hydrofluoric acid etch followed by UV-ozone
cleaning, (2) sapphire prepared with UV-ozone cleaning, and (3)
UV-grade fused silica cleaned with acetone, ultrapure methanol,
and de-ionized water.  Surfaces (1) and (2) displayed forces that
were 3--10 times larger than the Casimir-Polder force and
displayed significant spatial inhomogeneity. Previous studies of a
BK7 surface had led us to believe that magnetic impurities
embedded during the polishing process were a problem, thus leading
us to try an aggressive surface preparation such as a hydrofluoric
acid etch.  However, with our current surfaces, spurious forces
appear to be primarily caused by spatially inhomogeneous electric
surface potentials.

The fused silica surface (3) displayed the smallest level of
spurious forces.  Nevertheless, even with this surface we were
forced to study multiple spatial locations in order to locate
suitable positions for measurements.  Suitable locations were
primarily identified by the criteria that $\delta_\gamma$ be less
than a certain predefined threshold value.  Approximately 40$\%$
of the surface regions studied displayed spatially inhomogeneous
forces.  The spatially inhomogeneous forces, identified with the
technique previously described, displayed peak values
$\sim$100$\%$ larger than the Casimir-Polder force.  Spatial
variations occurred on $\sim$50 $\mu\mbox{m}$ distance scales and
displayed $\sim$$100\%$ percent variations in strength. It is
possible that finer structure could be present, yet not
detectable.

Once a suitable region was identified, we performed the
experimental procedure previously outlined to measure $\gamma_x$.
In addition, a significant amount of data was concurrently taken
to put limits on spatially uniform electric and magnetic forces.
Table~I summarizes the limits from these systematic measurements.
The results of our measurement of the Casimir-Polder force from
surface (3) are shown in Fig.~\ref{fig:data}.

The first thing to note is that our measurement distances are deep
within the retarded, or Casimir-Polder, regime.  This is
highlighted by the profound disagreement between the measured
force and the extrapolation of the van der Waals-London force to
this distance regime, as shown in Fig.~\ref{fig:data}.  We do, on
the other hand, see good agreement with the predicted
Casimir-Polder force from our fused silica surface. Unfortunately,
we currently do not have the experimental resolution to discern
between the $T$ = 0 K Casimir-Polder force and the $T$ = 300 K
case. Repeating the measurement at an elevated temperature,
however, appears promising; see Fig.~\ref{fig:data} for the
prediction for $T$ = 600 K.  At this temperature, the predicted
$\gamma_x$ is larger than nearly all of our data; thus a
measurement repeated at this temperature should yield a
significantly larger signal. Additionally, observation of a direct
correlation between a change in temperature and a corresponding
increase the Casimir-Polder force would clearly demonstrate the
thermal regime of the Casimir-Polder force.

\begin{figure}
\leavevmode
\epsfxsize=3.25in 
\epsffile{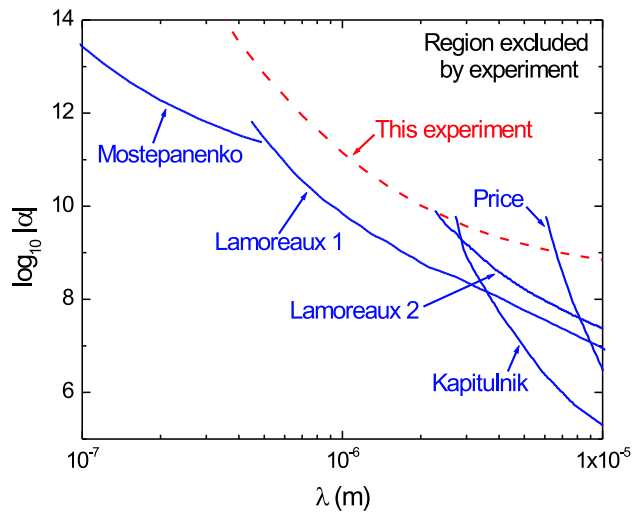} \caption{\label{fig:gravity} (Color
online) Current short-range Yukawa-type force limits in the
0.1--10 $\mu\mbox{m}$ range.  The limits obtained from this
experiment are shown by the dashed line. The limits labeled
Mostepanenko, Kapitulnik, and Price are from \cite{decca2005},
\cite{chiaverini2003}, and \cite{long2003}, respectively.  The
limits labeled Lamoreaux (a) and (b), from \cite{bordag2001} and
\cite{long1999}, respectively, are from two different analyses of
the Lamoreaux experiment \cite{lamoreaux1997}.}
\end{figure}

If we assume the $T$ = 300 K Casimir-Polder theory is correct,
then the data in Fig.~\ref{fig:data} can additionally be used to
put limits on short-range scalar-scalar Yukawa potentials of the
type~\cite{dimopoulos2003}

\begin{equation}
U_{Yuk} = - \int dV \frac{G m \rho}{r}(1 + \alpha e^{-r/\lambda}),
\end{equation}

\noindent where $G$ is the Newtonian constant of gravitation, $m$
is the mass of rubidium, $\rho$ is the density of the attracting
body (in our case the fused silica substrate), $r$ is the distance
from the rubidium atom to the volume element in the substrate,
$\alpha$ and $\lambda$ parametrize the short-range Yukawa force,
and the volume integral is performed over the the fused silica
substrate.

For each value of $\lambda$, we increase a hypothetical $\alpha$
until the value of the $\gamma_x$ predicted due to the
hypothetical Yukawa force plus the predicted $T$ = 300 K
Casimir-Polder force is excluded at the 95$\%$ level by the data
as shifted by worst-case assumptions on the systematics (for this
analysis uncertainties in spurious magnetic forces dominate).  The
limit obtained in this manner is plotted in Fig.~\ref{fig:gravity}
with the current experimental limits in this region. Redesigning
the experiment to optimize sensitivity to this signal could permit
over an order-of-magnitude improvement in the short-range force
sensitivity.   This improvement could be accomplished by, for
instance, using a material with significantly higher density or by
working over a surface where the condensate extends over two
materials of different density.

\section{Conclusion}

In summary, we have performed a precise measurement of the
Casimir-Polder force at a significantly larger atom-surface
separation than has previously been achieved.  At this large
atom-surface separation, effects due to thermal blackbody photons
become important, and extension of these measurements to
temperatures $\sim$300 K above room temperature should allow a
clear detection of this effect.  Additionally, future experiments
performed in nonequilibrium thermal conditions, such as holding
the surface at 600 K with the surroundings at 300 K, are predicted
to observe significant deviations from the equilibrium thermal
Casimir-Polder case~\cite{stringari2005}.  Study of the
Casimir-Polder force in such nonequilibrium situations will
hopefully permit a better understanding of this often nonintuitive
force.

Finally, this experiment has demonstrated promising short-range
force sensitivity that could provide limits on Yukawa-type forces
utilizing a significantly different measurement type--i.e.,
atom-bulk vs bulk-bulk.  Future work with ultracold atoms near
surfaces utilizing this technique and promising
atom-interferometry techniques currently being developed, should
enable better limits to be set in the 1--10 $\mu\mbox{m}$ regime.

\section{Acknowledgments}

We acknowledge useful conversations with Mauro Antezza, Lev
Pitaevskii, and Sandro Stringari as well as members of the JILA
BEC Collaboration.  We would additionally like to thank Simon
Kaplan for providing references for the optical properties of our
substrate materials.  This work was supported by grants from the
NSF and NIST and is based upon work supported under the NSF.


\begin{thebibliography}{10}

\bibitem[*]{SFU}
Present address: Simon Fraser University, Department of Physics,
8888 University Drive, Burnaby, British Columbia, Canada V5A 1S6.

\bibitem[\dag]{qpdNIST}
Quantum Physics Division, National Institute of Standards and Technology.

\bibitem{casimir1948}       
H.~B.~G. Casimir and P. Polder, Phys. Rev. {\bf 73}, 360 (1948).

\bibitem{hinds1993}         
C.~I. Sukenik, M.~G. Boshier, D. Cho, V. Sandoghdar, and E.~A. Hinds, Phys. Rev. Lett. {\bf 70}, 560 (1993).

\bibitem{lamoreaux1997}     
S.~K. Lamoreaux, Phys. Rev. Lett. {\bf 78}, 5 (1997).


\bibitem{aspect1996}        
A. Landragin, J.~Y. Courtois, G. Labeyrie, N. Vansteenkiste, C.~I. Westbrook, and A. Aspect, Phys. Rev. Lett. {\bf 77}, 1464 (1996).

\bibitem{shimizu2001}       
F. Shimizu, Phys. Rev. Lett. {\bf 86}, 987 (2001).

\bibitem{dekieviet2003}     
V. Druzhinina and M. DeKieviet, Phys. Rev. Lett. {\bf 91}, 193202 (2003).

\bibitem{vuletic2004}       
Y.~J. Lin, I. Teper, C. Chin, and V. Vuletic, Phys. Rev. Lett. {\bf 92}, 050404 (2004).

\bibitem{ketterle2004}      
T.~A. Pasquini, Y. Shin, C. Sanner, M. Saba, A. Schirotzek, D.~E. Pritchard, and W. Ketterle, Phys. Rev. Lett. {\bf 93}, 223201 (2004).

\bibitem{shimizu2005}       
H. Oberst, Y. Tashiro, K. Shimizu, and F. Shimizu, Phys. Rev. A {\bf 71}, 052901 (2005).

\bibitem{inguscio2004}      
G. Roati, E. de Mirandes, F. Ferlaino, H. Ott, G. Modugno, and M. Inguscio, Phys. Rev. Lett. {\bf 92}, 230402 (2004).

\bibitem{cornell2005}       
Y.-J. Wang, D.~Z. Anderson, V.~M. Bright, E.~A. Cornell, Q. Diot, T. Kishimoto, M. Prentiss, R.~A. Saravanan, S.~R. Segal, and S. Wu, Phys. Rev. Lett. {\bf 94}, 090405 (2005).

\bibitem{ketterle2005}      
M. Saba, T.~A. Pasquini, C. Sanner, Y. Shin, W. Ketterle, and D.~E. Pritchard, Science {\bf 307}, 1945 (2005).

\bibitem{stringari2004}     
M. Antezza, L. P. Pitaevskii, and S. Stringari, Phys. Rev. A {\bf 70}, 053619 (2004).

\bibitem{lewandowski2003}   
H.~J. Lewandowski, D.~M. Harber, D.~L. Whitaker, and E.~A. Cornell, J. Low Temp. Phys. {\bf 132}, 309 (2003).

\bibitem{harber2003}        
D.~M. Harber, J.~M. McGuirk, J.~M. Obrecht, and E.~A. Cornell, J. Low Temp. Phys. {\bf 133}, 229 (2003).

\bibitem{mcguirk2004}       
J.~M. McGuirk, D.~M. Harber, J.~M. Obrecht, and E.~A. Cornell, Phys. Rev. A {\bf 69}, 062905 (2004).

\bibitem{schmied2003}       
S. Schneider, A. Kasper, C. vom Hagen, M. Bartenstein, B. Engeser, T. Schumm, I. Bar-Joseph, R. Folman, L. Feenstra, and J. Schmiedmayer, Phys. Rev. A {\bf 67}, 023612 (2003).

\bibitem{threshold}         
The threshold limit for $\delta_\gamma$ was defined as the statistical error in $\gamma_x^{cm}$.  This definition, although somewhat arbitrary, carries significance because if $\delta_\gamma$ is much larger than the statistical error in $\gamma_x^{cm}$, then there is certainly a significant spatially inhomogeneous force present.

\bibitem{adsorbates}        
This is not such an unlikely situation.  In our previous work studying alkali adsorbates~\cite{mcguirk2004}, we measured electric fields from alkali atoms adsorbed onto surfaces.  The spatial distribution of atoms that adsorb onto the surface from the condensate will reflect the dimensions of the condensate and thus essentially form an elongated stripe of electrical dipoles collinear to the axis of the condensate.

\bibitem{bdot}              
$\partial \vec{B}/\partial t$ due to the charging and discharging of the plates is extremely small and can be safely neglected.

\bibitem{xygradient}        
$\vec{\nabla} \cdot \vec{B}=0$ forces the $\hat{x}$ and $\hat{y}$ magnetic field gradients to be equal at the center of the trap.

\bibitem{edot}              
$\vec{\nabla} \times \vec{B} \simeq 0$ because we are considering static fields in free space.

\bibitem{jarvis1977}        
J. Baker-Jarvis, R.~G. Geyer, J.~H. Grosvenor, Jr., M.~D. Janezic, C.~A. Jones, B. Riddle, C.~M. Weil, J. Krupka, IEEE Trans. Dielectric Electric. Insul. {\bf 5}, 571 (1998).

\bibitem{static-approx}     
For calculation of the Casimir-Polder force we utilize the static approximation of \cite{stringari2004} that relies only on the DC dielectric constant of the substrate.  The static approximation tends to slightly {\it overestimate} the force in the low-temperature limit.  So, for the smaller trap-center to surface separations, the T = 0 K prediction using the full theory would be slightly further below the T = 300 K prediction.  The size of this correction for T = 300 K is not enough to shift the limit curve in Fig.~\ref{fig:gravity}.

\bibitem{dimopoulos2003}    
S. Dimopoulos and A.~A. Geraci, Phys. Rev. D {\bf 68}, 124021 (2003).

\bibitem{decca2005}         
R.~S. Decca, D. Lopez, E. Fischbach, G.~L. Klimchitskaya, D.~E.
Krause, V. M. Mostepanenko, Ann. Phys. (N.Y.) {\bf 318}, 37
(2005).

\bibitem{chiaverini2003}    
J. Chiaverini, S.~J. Smullin, A.~A. Geraci, D.~M. Weld, and A. Kapitulnik, Phys. Rev. Lett. {\bf 90}, 151101 (2003).

\bibitem{long2003}          
J.~C. Long, H.~W. Chan, A.~B. Churnside, E.~A. Gulbis, M.~C.~M. Varney, and J.~C. Price, Nature {\bf 421}, 922 (2003).

\bibitem{bordag2001}        
M. Bordag, U. Mohideen, and V.~M. Mostepanenko, Phys. Rep. {\bf 353}, 1 (2001).

\bibitem{long1999}          
J.~C. Long, H.~W. Chan, and J.~C. Price, Nucl. Phys. B {\bf 539} (1999) 23.

\bibitem{stringari2005}     
M. Antezza, L.~P. Pitaevskii, and S. Stringari, Phys. Rev. Lett.
{\bf 95}, 113202 (2005).

\end{thebibliography}
\end{document}